# Runaway electron modelling in the self-consistent core European Transport Simulator, ETS

**Gergo I. Pokol[1], Soma Olasz[1], Boglarka Erdos[1], Gergely Papp[2], Matyas Aradi[3], Mathias Hoppe[4], Thomas Johnson[5], Jorge Ferreira[6], David Coster[2], Yves Peysson[7], Joan Decker[8], Par Strand[4], Dimitriy Yadikin[4], Denis Kalupin[9] and the EUROfusion-IM Team[*]**

[1] NTI, Budapest University of Technology and Economics, Budapest, Hungary
[2] Max-Planck Institute for Plasma Physics, Garching, Germany
[3] Graz University of Technology, Fusion@ÖAW, Graz, Austria
[4] Chalmers University of Technology, Gothenburg, Sweden
[5] KTH Royal Institute of Technology, Stockholm, Sweden
[6] IPFN, IST, Universidade de Lisboa, Lisboa, Portugal
[7] CEA, IRFM, Saint-Paul-lez-Durance, France
[8] EPFL, Swiss Plasma Center, Lausanne, Switzerland
[9] EUROfusion Programme Management Unit, Garching, Germany

E-mail: pokol@reak.bme.hu

## Abstract

Relativistic runaway electrons are a major concern in tokamaks. Albeit significant theoretical development had been undertaken in the recent decades, we still miss a self-consistent simulator that could simultaneously capture all aspects of this phenomenon. The European framework for Integrated Modelling (EU-IM), facilitates the integration of different plasma simulation tools by providing a standard data structure for communication that enables relatively easy integration of different physics codes. A three-level modelling approach was adopted for runaway electron simulations within the EU-IM. Recently, a number of runaway electron modelling modules have been integrated into this framework. The first level of modelling (Runaway Indicator) is limited to the indication if runaway electron generation is possible or likely. The second level (Runaway Fluid) adopts an approach similar to e.g. the GO code, using analytical formulas to estimate changes in the runaway electron current density. The third level is based on the solution of the electron kinetics. One such code is LUKE that can handle the toroidicity-induced effects by solving the bounce-averaged Fokker-Planck equation. Another approach is used in NORSE, which features a fully nonlinear collision operator that makes it capable of simulating major changes in the electron distribution, for example slide-away. Both codes handle the effect of radiation on the runaway distribution. These runaway-electron modelling codes are in different stages of integration into the EU-IM infrastructure, and into the European Transport Simulator (ETS), which is a fully capable modular 1.5D core transport simulator. ETS with Runaway Fluid was benchmarked to the GO code implementing similar physics. Coherent integration of kinetic solvers requires more effort on the coupling, especially regarding the definition of the boundary between runaway and thermal populations, and on consistent calculation of resistivity. Some of these issues are discussed.

---

[*] http://euro-fusionscipub.org/eu-im

## 1. Introduction

Relativistic runaway electrons are of major concern in tokamaks. They may significantly affect discharge properties mainly in low density regimes and during the startup phase, and a high-current runaway electron beam formed in disruptions is one of the most critical problems of reactor-size tokamak-type devices [1,2]. Several tools, addressing various aspects of the problem, have been developed in the recent decades, like the bounce-averaged kinetic solver LUKE [3,4], the kinetic solver NORSE with non-linear collision operator [5], or test particle following in non-linear MHD simulations [6] (for a review, see [7]). There was also work done using test-particles to study confinement and mitigation of runaways in 3-D toroidal plasmas [8-11]. However, we still miss a self-consistent simulation tool that could simultaneously capture all aspects of this phenomenon. This paper presents some integration steps towards the development of such modelling capabilities.

Integration of different plasma simulation tools, often working on different time and spatial scales and handling varying levels of details, can be difficult. The EUROfusion Code Development for integrated modelling project (WPCD) facilitates this by providing an Integrated Modelling framework (EU-IM) [12], and a standard data structure for communication [13] that enables relatively easy integration of different physics codes into a complex scientific workflow. Interchangeability of the physics modules is a main feature that allows easy benchmarking, which is extremely useful for the verification of the workflows, and also allows the exploration of the range of validity of each code. EU-IM has adopted a graphical workflow interface, Kepler [14], in which components are called "actors". EU-IM hosts a number of workflows for different applications [15]. The most sophisticated EU-IM workflow is the European Transport Simulator (ETS), which is a fully capable modular 1.5D core transport simulator [16,17]. It features consistent treatment of different plasma species and kinetic modelling of non-thermal ions and electrons originating in heating and current drive, which makes it possible to use already existing data bundle and workflow structure for runaway electron modelling. The ITER Integrated Modelling and Analysis Suite (IMAS) has been developed along the same concepts [18,19], and adapting the ETS, together with its components, to IMAS is in progress, but results shown in the present paper were achieved in the EU-IM framework.

Section 2 describes the step-by-step approach to integrate runaway electron modelling capabilities into the EU-IM framework. First results of simple runaway models integrated into ETS are described in Section 3.1, while progress and known issues of the integration of kinetic models are reported in Section 3.2. The paper is concluded in Section 4.

## 2. Step-by-step approach of model integration into EU-IM

The integration process of simulation codes to the EU-IM framework consists of steps like converting the inputs and outputs to standard data structures called Consistent Physical Objects (CPOs) [13], providing automated test and build procedures, user documentation, and finally producing an "actor" that can be pulled into a Kepler workflow. In order to ensure maximum effectiveness of the effort, a three-level modelling approach was adopted to runaway electron simulation within the EU-IM [20].

The first level of modelling (*Runaway Indicator*) is limited to the indication if runaway electron generation is possible or likely. *Runaway Indicator* has two functions: It generates a warning message if the E parallel electric field is higher than the Ec critical field for runaway generation [21] anywhere inside the x = r/a = 0.95 normalized minor radius. It gives a second warning if the toroidal electric field in this region is expected to produce a non-negligible runaway current according to the widely used Connor & Hastie primary generation formula (67) [21]. This is expected to discriminate false indications of flattop runaways [22]. The default value for the threshold is $10^{12}$ s$^{-1}$m$^{-3}$, that is expected to produce 1 kA of runaway current in about 10 s, but it is to be customized based on threshold for detecting the effect of the runaway electrons and discharge duration.

The second level (*Runaway Fluid*) adopts a similar approach to the GO code [23], using analytical formulas to estimate changes in the runaway electron current density, and by assuming the parallel velocity of all runaway electrons to be close-to the speed of light in the calculation of the runaway current density. For primary generation it takes the Dreicer generation into account by either the most general formula (63) of Connor and Hastie [17] or formula (66) valid for high E/Ec normalized electric field, or even the simplest formula (67) constrained to relatively low temperatures, but providing a systematic overestimation of runaway generation in the whole domain. For realistic aspect ratio tokamaks a correction factor for the effect of toroidicity can optionally be applied as suggested by Nilsson et.al. [24]. *Runaway Fluid* uses the classical formula for avalanche generation by Rosenbluth and Putvinski [25]. Their formula can optionally be modified by a Ea threshold at low electric field obtained using a momentum-conserving approach to the knock-on collisions and approximated by formula (8) of the paper by Aleynikov et.al. [26]. Studies with LUKE [24] showed that the avalanche growth rate can also be significantly reduced at toroidal magnetic surfaces with high mirror ratio due to the trapping of the high energy electrons generated in the knock-on collisions. For this purpose formula (A.4) of the paper by Nilsson et.al. [24] is implemented in *Runaway Fluid*. Combinations of formulae can be customized by code parameters. The implemented correction factors are progressively updated based on most recent results on the details of relevant collision processes [27,28,29].

us to set the desired density, temperature and electric field and study the time evolution of the runaway electron population.

The third level is based on the solution of the electron kinetics. This is performed via a code such as LUKE [3,4] that can handle the toroidicity-induced effects by solving the bounce-averaged Fokker-Planck equation. Another approach is used in NORSE [5], which features a fully nonlinear collision operator that makes it capable of simulating major changes in the electron distribution, for example slide-away [30]. Both codes handle the effect of radiation on the runaway distribution, that was recently shown to be an important factor [22,31]. Both LUKE and NORSE are now available for EU-IM, while interfacing and producing the actors is in progress. The second level *Runaway Fluid* code is already interfaced to EU-IM data structure such that it is exchangeable with the kinetic codes.

## 3. Runaway models in the European Transport Simulator

Having finished the integration and testing of runaway-electron modelling codes in the EU-IM infrastructure, we could proceed with the integration into the ETS workflow [16,17]. *Runaway Indicator* is integrated into the *Instantaneous Events* module of ETS, and by default it runs in every time step. Most of the time it indicates that there is no possibility of runaway electron production, thus the simulation results are valid without any further runaway electron modelling.

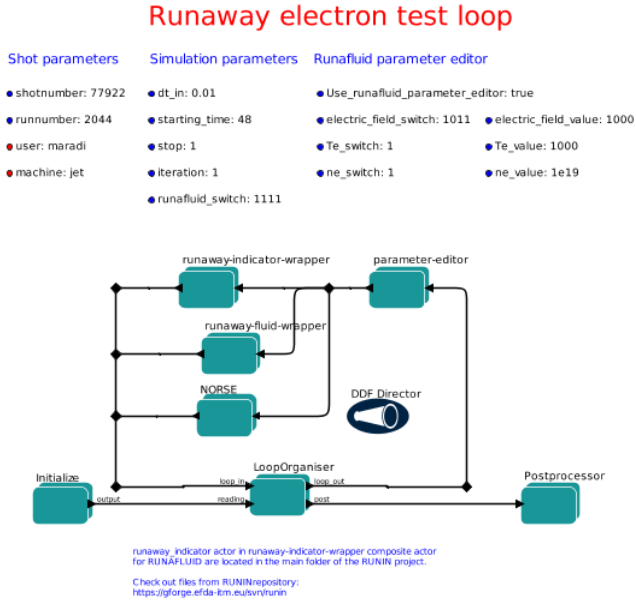

*FIG. 1. Runaway electron test Kepler workflow in EU-IM.*

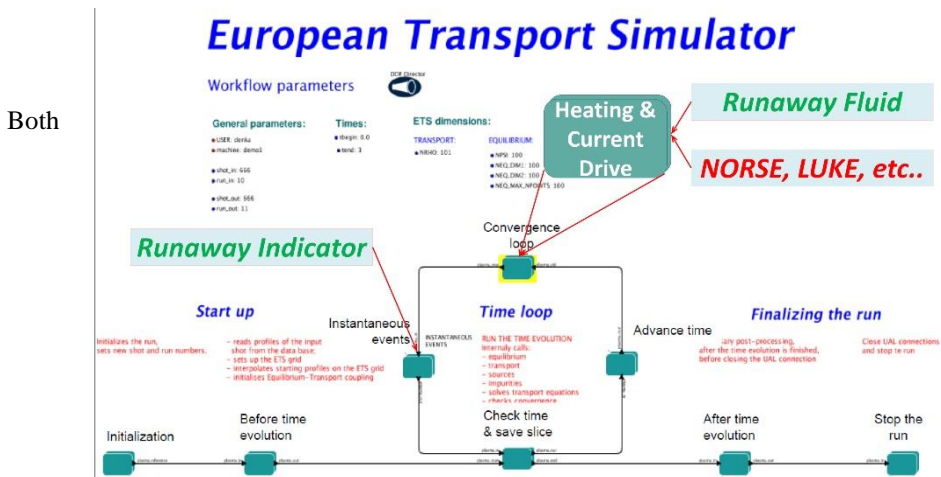

*Runaway Indicator* and *Runaway Fluid* have been tested on many levels from unit testing to integration testing utilizing a custom made test workflow illustrated in Fig. 1. This enabled

*FIG. 2. The place of runaway electron actors within ETS. The green boxes represent different sub-workflows (composite actors), having their own internal structure including different physics actors.*

The more sophisticated runaway electron models providing estimates of the non-inductive runaway current and other properties were integrated into the *Heating & Current Drive* workflow, which itself is part of the *Convergence Loop* of ETS, as shown in Fig. 2.

By default, actual runaway electron modelling is switched off in ETS, otherwise it allows exact specification of models to be used when enabled by an expert user. This ensures a self-consistent evolution of the runaway electron distribution and macroscopic plasma properties such as the toroidal electric field.

### 3.1 Integration of fluid-like models

*Runaway Fluid* needs the thermal plasma properties and the electric field as an input, and provides an estimate of the runaway electron current as an output. This non-inductive current can then be taken into account when solving the electric field diffusion equation by ETS. ETS with *Runaway Fluid* was benchmarked to the GO code [23] implementing similar physics [32]. The difficulty with this task is that the simple runaway models of GO were shown to be relevant for large electric field cases with self-consistent electric field diffusion, like disruptions [23], but they are not valid for quasi-steady state conditions, which is the usual operation scenario for ETS. For the purpose of the benchmark, a new actor was implemented in ETS that produces an energy sink for electrons and ions with a power proportional to the energy content of the corresponding population, thus producing an exponential drop in temperature with a specified td decay time. It also has a feature to smoothly stop the temperature drop at a specified Tmin minimum temperature. Having introduced this drastic change in energy content on the timescale of milliseconds, we switched off all other transport models and sources. A common choice for the boundary condition on the current diffusion equation was the perfectly conducting wall just at the plasma boundary. The benchmark was performed with td=0.5 ms and Tmin=15 eV starting from an ASDEX-Upgrade like scenario as initial condition. Qualitative and order of magnitude correspondence was found between GO and ETS, but there was also a significant difference in the evolution of the electric field and as a result there was a factor of 2 difference in the runaway current as shown in Fig. 3. This can probably be explained by the different assumptions on magnetic geometry.

The validity of the analytical formulas in *Runaway Fluid* is limited. To some degree, this range of validity is extended by introducing correction factors detailed in Section 2, some of which have been found to have a significant effect on runaway electron production in minor disruption-like transients. The bottom plots of Fig. 3. show the results employing all the correction factors.

Integration of *Runaway Fluid* already raised some numerical issues that needed to be handled. A numerical instability was observed when trying to calculate with too large time step resulting from the explicit way of handling the runaway electron production: An insufficiently large time step could cause an overshoot of runaway electron current, which causes a large electric field to appear in the opposite direction. In the next time step this opposite direction electric field would cause an overshoot in the opposite direction, resulting in an even larger electric field in the original direction, inducing an alternatingly diverging process. This was eliminated by implementing a numerical check that in any time step just a small fraction of electrons can become runaway, which resulted in several hundreds of time steps within the duration of the thermal quench.

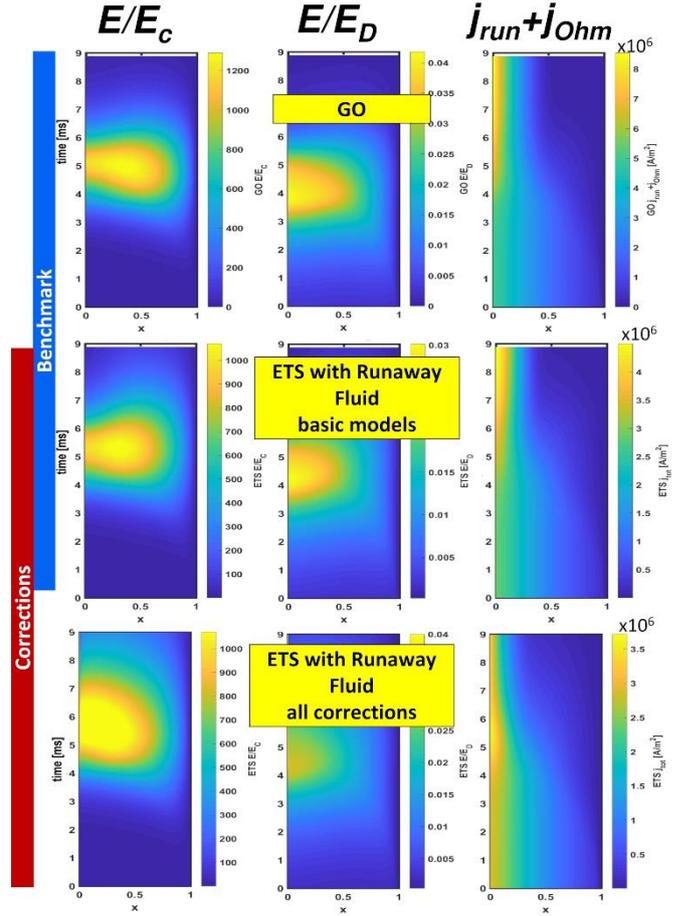

FIG. 3. Results of the ETS with Runaway Fluid benchmark: evolution of the 1D electric field (E/Ec(x)) and the total current density (jtot(x)) profiles for GO (top), the simplest classic models for runaway electron generation in Runaway Fluid (middle) compared to the case of applying various kinetic corrections (bottom).

A more subtle issue is that the presence of runaway electrons compromises the calculation of plasma resistivity. At present it is assumed that some conductivity is carried by the bulk electrons, and some conductivity is taken care of by the runaway actor through the runaway electron growth rates. This is a good model as long as only a small number of runaway electrons is present, but having a severe deformation of the distribution function leads to the kinetic nature of the problem.

*Runaway Fluid* supports a number of corrections to the original runaway electron growth rates, but even with these extensions, the *Runaway Fluid* approach does not provide reliable modelling for near critical electric field cases. Yet, it can still be used to extend the validity of ETS to scenarios having just a little bit of runaways: The modelling can be first run by neglecting the effect of runaways, which is the default setting. Then if the *Runaway Indicator* gave warnings, the model can be re-run with Runaway Fluid set to use the growth rates without (most of) the corrections. In this case, *Runaway Fluid* gives a conservative over-estimation of the runaway current. If the comparison of the two cases shows no significant difference, the user can be sure that runaway electrons are not a significant factor in the studied simulation scenario.

### 3.2 Integration of kinetic models

Both LUKE [3,4] and NORSE [5] are in some stage of integration into EU-IM, already. However, coherent integration of kinetic solvers – that produce the evolution of the full electron distribution – into a 1.5D transport code requires more effort on the coupling, especially regarding the definition of the boundary between runaway and thermal populations, and on consistent calculation of resistivity.

Actually, defining runaway electrons as the part of the electron distribution function featuring continuous acceleration up to relativistic energy does not suit our needs. If – for example – the electric field decreases below the critical field, continuous acceleration stops, yet the behaviour of the highly relativistic electron population does not change suddenly. Possibly the best approach is to define a fixed boundary between bulk and supra-thermal electrons, and stick to it, as proposed by Papp et al. [33]. The exact location of the boundary should not affect the result very sensitively, but a check is certainly a good idea to implement. Whenever it is not possible to define such boundary, because of e.g. slide-away distortion of the whole distribution function [5], separation between bulk and runaway populations really breaks down, our way of handling runaway electrons in ETS becomes invalid.

The output of both kinetic codes is the electron distribution function, as opposed to *Runaway Fluid* providing a sole runaway electron density, which can thus be susceptible to more kinds of instabilities. Outputs should therefore always be checked for non-physical oscillations in phase space and negative distribution function values before they are integrated in velocity space to produce the non-inductive current for the electric field diffusion equation in ETS.

Some kinetic solvers have their own model for radial runaway electron transport implemented (like LUKE [4]), but it typically only describes diffusion with constant diffusion coefficients, which was shown to be insufficient for at least a subset of relevant cases [34]. Monte Carlo orbit following of runaway electrons at every time step is too expensive computationally, but it has been showed that an advection–diffusion model with coefficients fit from orbit-following calculations might be sufficient to describe cross-field runaway electron transport in perturbed magnetic fields [35]. Integration of such capability could account for the dominant loss mechanism of runaway electrons in perturbed magnetic fields.

## 4. Conclusions

Developing the runaway electron modelling capability within the EU-IM / IMAS framework and more specifically the European Transport Simulator, ETS, is an ongoing effort. The first two stages are already operational, featuring the integration of *Runaway Indicator* and *Runaway Fluid* actors. These actors can detect the possible generation of runaway electrons and give a conservative estimate of their effect, which already extended somewhat the range of applicability of ETS. These actors have been tested in custom developed workflows and ETS with *Runaway Fluid* has been benchmarked against GO [23,32]. However, integration of proper kinetic Fokker-Planck solvers is necessary for realistic modelling of scenarios with runaway electrons. LUKE [3,4] and NORSE [5] have been chosen for integration, as they have complementary capabilities, one describing the toroidicity effects relevant mostly for steady-state cases, the other featuring non-linear collision operator useful in high electric field cases, respectively. Kinetic solvers for runaway electrons are being integrated parallel to *Runaway Fluid*, which ensures the possibility of easy future comparison of results and benchmarking. There are operation scenarios and stages of disruptions, when LUKE and NORSE might not contain sufficient physics. For these cases step 4 of code integration is to be devised and implemented.

## Acknowledgements

This work has been carried out within the framework of the EUROfusion Consortium and has received funding from the Euratom research and training programme 2014-2018 and 2019-2020 under grant agreement No 633053. The views and opinions expressed herein do not necessarily reflect those of the European Commission.